\documentclass[12pt]{article}
\usepackage{amssymb,amsmath,graphicx,latexsym,cite}
\usepackage[usenames]{color}
\begin{document} 
\title{\textbf{The accidental Higgs}} 
\author{B.~Holdom\thanks{bob.holdom@utoronto.ca}\\
\emph{\small Department of Physics, University of Toronto}\\[-1ex]
\emph{\small Toronto, Ontario M5S1A7, Canada}}
\date{}
\maketitle
\begin{abstract}
We suggest that the Higgs boson is a light composite state that does not emerge from TeV scale strong dynamics for any generic reason, such as when it is pseudo-Goldstone boson. Instead, a state that is Higgs-like and fairly decoupled from heavier states may simply be a reflection of very particular strong dynamics, with properties quite distinct from more familiar large-$N_c$ type gauge dynamics. We elaborate on this picture in the context of a strongly interacting fourth family and an effective 4-Higgs-doublet model. The origin of a decoupling limit and the corrections to it are discussed. 
\end{abstract}
The discovery of the 126 GeV mass Higgs would appear to resolve the question of the origin of electroweak symmetry breaking, especially given the simplicity of the standard model description (SM) and the lack of any evidence of additional physics. But nagging issues of naturalness and the long list of parameters of the standard model suggest that we have not yet reached the end of the story. If the Higgs description emerges as only an effective low energy description then it is with the ultraviolet completion that the story can continue.

But ultraviolet completions that maintain the local Higgs description on scales at least an order of magnitude above the TeV scale are now facing their own issues of naturalness as well as a general lack of simplicity. Of course the reason to push the ultraviolet completions to higher scales is to avoid effects of the new physics that perhaps should already have been seen. But we view the exact nature of the new effects to be a model dependent question, so that the generic estimates of the effects may be substantially modified by the structure of a particular theory or by difficult to calculate effects of strong interactions. With this in mind we feel that it remains worthwhile to consider the possibility that the Higgs description breaks down at no more than a few times the TeV scale, since it is in this case that issues of naturalness are most simply resolved. It may appear that nature has conspired a little to keep this nearby physics hidden from us, but we won't know by how much unless we study these theories further.

From the viewpoint of simplicity and economy, the main advantage of an ultraviolet completion at a TeV is that the fundamental matter degrees of freedom can be standard chiral fermions. Rather than exotically charged fermions, a sequential extension of the known fermions, a fourth family, can be considered. New fermions of a fourth family have masses that are bounded from above, $\lesssim 1$ TeV, due to the fact that their masses would contribute to $W$ and $Z$ masses. The direct search for the heavy quarks of a fourth family have not yet saturated this bound, but the current lower limits on their masses do put the nature of new interactions involving the fourth family firmly in the strongly interacting regime.

In the case of a strongly interacting fourth family it is the condensates of the heavy quarks, the $t'$ and the $b'$, that are likely the primary origin of electroweak symmetry breaking. We shall find that it is also important to include the effects of the $t$ and the $\tau'$, and so we include all four fermions in the set of heavy fermions we consider. They all contribute to the loop induced $gg$ and/or $\gamma\gamma$ couplings of the light states of interest while the fourth neutrino $\nu_\tau'$ does not and so we choose to neglect it. A neutrino condensate does contribute to electroweak symmetry breaking but the error we make by neglecting it should be small. In addition if this neutrino mass is of the Majorana type then the mixing of the associated scalar mode with the other scalar modes should be suppressed. There are various contributions of both signs that the heavy fermions make to the $S$ and $T$ parameters, but a fourth family cannot as yet be ruled out solely by these precision measurements \cite{Holdom:2006mr,Kribs:2007nz}.

The discovery of the Higgs-like 126 GeV state presents some serious hurdles for any theory of dynamical electroweak symmetry breaking. There are basically three questions. (1) How can the strongly interacting theory at a TeV have a light scalar in its mass spectrum? (2) Why should this light scalar resemble a fluctuation around the vacuum expectation value (vev) of an electroweak scalar doublet, as indicated by its observed couplings to $W$ and $Z$? (3) How can it be that the couplings of the scalar to the heavy fermions are such that the induced loop couplings to $gg$ and $\gamma\gamma$ also resemble that of the Higgs boson?

We first comment on the second question. We are interested in the condensates that can develop for the four scalar electroweak doublets $\bar{t}_R q_L$, $\bar{t}'_R q'_L$, $\bar{b}'_R q'_L$, $\bar{\tau}'_R \ell'_L$. The fluctuations around these condensates include the neutral and charged Goldstone bosons. They may also include other rather light states, at least lighter than twice the heavy quark mass. The local fermion condensates are one manifestation of the symmetry breaking, but a better representation of the order parameters is provided by the momentum dependent dynamically generated fermion mass functions. Then differing fluctuations around these mass functions will also be characterized by their differing momentum dependence, that is by their form factors in momentum space.

The main question is the nature of the form factor of the lightest neutral scalar fluctuation (one for each flavor). The point is that this form factor may have a momentum dependence that is similar to that of the mass function.\footnote{As shall be discussed elsewhere \cite{cat}, the integral equations that determine the mass function and the form factor only differ by terms that become important in the infrared, that is by terms that implement an effective infrared cutoff in the respective integral equation. Thus the solutions will be similar for momenta above this cutoff.} That is the lowest lying scalar fluctuation of the mass function is close to being a fluctuating multiplicative factor times the mass function. This means that the attachment of a low momentum scalar to a fermion loop hardly changes the value of the loop, or in other words the amplitude for an additional scalar is close to being $\sigma(x)/v$ times the original amplitude.\footnote{In more detail one needs to distinguish $v=f_\pi$ and $f_\sigma$, but again the (small) difference is due to terms that implement an effective infrared cutoff in the respective loop integrals.} This is a property of a linear sigma model description, in which the action is a function of $v+\sigma(x)$. $SU(2)\times U(1)$ symmetry must also be manifest and so the approximate low energy description must be in terms of electroweak scalar doublets fluctuating about their vevs.\footnote{The authors of \cite{Belyaev:2013ida} argue that this resemblance even extends to the sigma resonance of QCD. They obtained the $\sigma\pi\pi$ coupling in 2-flavor QCD from the $I=0$ and $J=0$ partial-wave projection of the elastic $\pi\pi$ scattering amplitude at the $\sigma$ pole \cite{GarciaMartin:2011jx}. The value of the $\sigma\pi\pi$ coupling so obtained agrees very well with the linear sigma model prediction. This coupling is the analog of the Higgs coupling to $WW$ and $ZZ$.}
 
We thus pursue an effective scalar field description in which we have four electroweak doublets,
\begin{align}
\Phi_i=\left(\begin{array}{c}\phi_i^+ \\ (v_i+\sigma_i+i\eta_i)/\sqrt{2}\end{array}\right),\quad i=t,t',b',\tau'
\label{e4}\end{align}
 with $\sum_i v_i^2=v^2$. This effective theory need only be well behaved for field values and field momenta less than a compositeness scale, at most a few times a TeV. The four doublets have hypercharge $+1$, and $\tilde{\Phi}_t$, $\tilde{\Phi}_{t'}$, $\Phi_{b'}$, $\Phi_{\tau'}$ (where  $\tilde{\Phi}\equiv i\tau_2\Phi^*$) have the quantum numbers of the fermion bilinears $\bar{t}_R q_L$, $\bar{t}'_R q'_L$, $\bar{b}'_R q'_L$, $\bar{\tau}'_R \ell'_L$. There is a rough proportionality between the vevs $v_i$ and the underlying dynamical fermion masses $m_i$. This is seen in the one loop contribution to the $W$ and $Z$ masses which can be written approximately in the form
\begin{align}
v_i^2=\frac{n_i m_i^2}{4\pi^2}\ln\frac{\Lambda_i}{m_i}
.\label{e1}\end{align}
$n_i=1$ or 3 is the color factor and $\Lambda_i$ characterizes the scale of significant falloff of the mass function. The large $t'$ and $b'$ masses should be similar and they are basically determined so that the correct $v$ emerges. Masses around 800 GeV would mean that they have roughly the same ratio to $v$ as the constituent quark masses have to $f_\pi$ in QCD. 

We see then that $v$ is well below the compositeness scale $\sim 2m_{q'}$, as is needed for self-consistency of the effective scalar description. This also means that Yukawa couplings are large, the Yukawa coupling of the $\Phi_i$ field to the $i$th heavy fermion is $\sqrt{2}m_i/v_i$. From (\ref{e1}) this gives a Yukawa coupling $\sim5$ for the heavy quarks; this is pushing into the unitarity bound but this is just a reflection of an underlying strongly interacting (and unitary) theory. Also if the Yukawa coupling was probed on scales of the order of the compositeness scale or larger, a damping form factor would become apparent.

The $\tau'$ mass is likely closer to the $t$ mass than to the $t'$ and $b'$ masses \cite{Holdom:1988yj}. Thus we shall be assuming a clear separation between the large vevs $v_{t'}$ and $v_{b'}$ and the smaller vevs $v_{\tau'}$ and $v_t$.  To be definite we shall set\footnote{The choice of $v_{\tau'}=v_t$ will simplify our discussion but it is not crucial for our results.} 
\begin{align}
v'\equiv v_{t'}=v_{b'}=t_\chi v_{t}=t_\chi v_{\tau'}
\end{align}
with $t_\chi\equiv\tan\chi\sim m_{q'}/m_t\sim5$ so that $v'^2=\frac{1}{2}v^2\sin^2\chi\approx\frac{1}{2}v^2$. In the following this will lead to an expansion in powers of $1/t_\chi$.

We label the four neutral scalar mass eigenstates $h_1$, $h_2$, $h_3$, $h_4$ ordered from small to large mass. Of most interest is the lightest state $h_1=\sum_i s_i \sigma_i$ with $\sum_i s_i^2=1$. With standard kinetic terms for the $\Phi_i$ the coupling of $h_1$ to $WW$ and $ZZ$ is proportional to $v^{-1}\sum_{i=1}^4 s_i v_i$. The maximum value of this is unity, the value for the SM Higgs boson, which occurs for $s_i=v_i/v$. In our case this is $[s_t, s_{t'}, s_{b'}, s_{\tau'}]=\frac{\sin\chi}{\sqrt{2}}[1/t_\chi,1,1,1/t_\chi]$. $h_1$ has the Yukawa couplings $h_1\sum_i(s_i m_i/v_i)\bar{\psi}_i\psi_i$ to the heavy fermions, and these also take values expected of a Higgs boson when $s_i=v_i/v$. In our framework there is no reason that it is precisely the $s_i=v_i/v$ combination that is a mass eigenstate, but data tells us that the 126 GeV state is not too far from it. 

We can now turn to the third question posed above. A fourth family yields additional loop contributions to Higgs couplings to $gg$ and $\gamma\gamma$ and these couplings are typically driven very far from SM couplings. At least this is true for the combination $s_i=v_i/v$. In particular the $h_1gg$ loop amplitude relative to the SM value in the heavy quark loop approximation is $v\sum_{\{t,t',b'\}} s_i/v_i$, which for $s_i=v_i/v$ is 3. The fermion loop contribution to the $h_1\gamma\gamma$ amplitude relative to the standard top loop contribution is $v(s_t/v_t+s_{t'}/v_{t'}+\frac{1}{4}s_{b'}/v_{b'}+\frac{3}{4}s_{\tau'}/v_{\tau'})$, which for $s_i=v_i/v$ would again be 3. But instead consider $[s_t, s_{t'}, s_{b'}, s_{\tau'}]=\frac{\sin\chi}{\sqrt{2}}[-1/t_\chi,1,1,1/t_\chi]$. In this case both of the previous amplitude factors take the value of unity.\footnote{If the $t'$, $b'$ and $\tau'$ contributions to these amplitudes were uniformly increased or decreased, perhaps due to some new strong interaction effect we have neglected, then a further shift in $s_t$ could again bring both amplitudes back to unity.} The $h_1$ coupling to $WW$ and $ZZ$ is then necessarily smaller, but it is only slightly smaller by a factor of $(1+1/t_\chi^2)^{-1}=\sin^2\chi$. Thus only a change of sign of the small top component of the $h_1$ field, $s_t\to-s_t$, brings its couplings dramatically closer to the standard values. In the following we shall be concerned with how this change of sign can arise through a study of the mass matrix for the scalars.

The $b$ quark and lighter fermions must also have Yukawa couplings that are induced by some underlying flavor physics. These Yukawa couplings can in principle involve all of the four scalar doublets $\Phi_i$. For example the $3\times3$ down-type quark mass matrix is $\sum_i v_i Y^{d}_i$ where the $Y^d_i$ are four Yukawa coupling matrices. Meanwhile the $h_1$ coupling matrix to the down-type quarks is $\sum_i s_i Y^{d}_i$. For the special case $s_i=v_i/v$ these two matrices are proportional and $h_1$ does not have flavor changing couplings. But when $s_t\to-s_t$ then there can be flavor changing couplings that are suppressed by ${\cal O}(1/t_\chi)$. The $h_1 \bar{b}b$ coupling could also receive a ${\cal O}(1/t_\chi)$ correction. The actual size of these effects is of course dependent on the form and relative sizes of the $Y^d_i$. The situation is similar for the charged leptons involving the Yukawa matrices $Y^e_i$, and if for example $Y^e_t$ was small compared to $Y^e_{\tau'}$ then the correction to the $h_1 \bar{\tau}\tau$ coupling may be smaller than ${\cal O}(1/t_\chi)$. For up-type quarks the difference is that the $t$ mass comes solely from $\Phi_t$ and the $h_1\bar{t}t$ coupling $s_t m_t/v_t=-m_t/v$ is negative. This sign has physical effects since it is relative to the still positive top mass.

There remains the first question posed above: why is there any light scalar at all? We expect the $(t',b')$ sector to display an approximate $SU(2)_L\times SU(2)_R$ symmetry and so it will be useful to use a notation that makes this explicit. We define the fields ${\cal U}=(\tilde{\Phi}_{t'}\;\Phi_{b'})$ and $\tilde{\cal U}=(i\tau_2){\cal U}^*(i\tau_2)^T=(\tilde{\Phi}_{b'}\;\Phi_{t'})$ (these $2\times2$ matrices are not constrained to be unitary) that both transform like ${\cal U}\to U_L{\cal U}U_R^\dagger$, under $SU(2)_L\times SU(2)_R$. Mass terms expressed in terms of these fields take the familiar forms (with ${\rm Tr}(\tilde{{\cal U}}^\dagger\tilde{{\cal U}})={\rm Tr}({\cal U}^\dagger{\cal U})$).
\begin{align}
&m_2^2{\rm Tr}({\cal U}^\dagger{\cal U})-\frac{1}{2}(m_3^2{\rm Tr}(\tilde{\cal U}^\dagger{\cal U})+{\rm h.c.})\nonumber\\
=\;&m_2^2(\Phi_{t'}^\dagger\Phi_{t'}+\Phi_{b'}^\dagger\Phi_{b'})-(m_3^2\Phi_{t'}^\dagger\Phi_{b'}+{\rm h.c.})
\label{e2}\end{align}
The mass mixing term plays an essential role in 2-Higgs-doublet models and it is clearly consistent with the $SU(2)_L\times SU(2)_R$ symmetry.

We can now construct the $SU(2)_L\times SU(2)_R$ symmetric quartic terms from the ${\cal U}$ and $\tilde{\cal U}$ fields. We have the one trace
\begin{align}
&\kappa_1{\rm Tr}({\cal U}^\dagger{\cal U}{\cal U}^\dagger{\cal U})+\kappa_{2}{\rm Tr}({\cal U}^\dagger{\cal U}\tilde{\cal U}^\dagger\tilde{\cal U})+\kappa_{3}{\rm Tr}({\cal U}^\dagger\tilde{\cal U}\tilde{\cal U}^\dagger\cal U)\nonumber\\
&+\lbrace{\kappa_{4}\over2}{\rm Tr}({\cal U}^\dagger{\cal U}\tilde{\cal U}^\dagger{\cal U})+{\kappa_{5}\over2}{\rm Tr}(\tilde{\cal U}^\dagger{\cal U}\tilde{\cal U}^\dagger{\cal U})+{\rm h.c.}\rbrace
,\end{align}
and the two trace terms
\begin{align}
&\hat{\kappa}_1{\rm Tr}({\cal U}^\dagger{\cal U}){\rm Tr}({\cal U}^\dagger{\cal U})+\hat{\kappa}_{2}{\rm Tr}({\cal U}^\dagger{\cal U}){\rm Tr}(\tilde{\cal U}^\dagger\tilde{\cal U})+\hat{\kappa}_{3}{\rm Tr}({\cal U}^\dagger\tilde{\cal U}){\rm Tr}(\tilde{\cal U}^\dagger{\cal U})\nonumber\\
&+\lbrace{\hat{\kappa}_{4}\over2}{\rm Tr}({\cal U}^\dagger{\cal U}){\rm Tr}(\tilde{\cal U}^\dagger{\cal U})+{\hat{\kappa}_{5}\over2}{\rm Tr}(\tilde{\cal U}^\dagger{\cal U}){\rm Tr}(\tilde{\cal U}^\dagger{\cal U})+{\rm h.c.}\rbrace
.\end{align}
More insertions of $\tilde{\cal U}$ do not produce new terms.

Let us consider one particular degree of freedom $\sigma(x)$ where
\begin{align}
{\cal U},\;\tilde{\cal U}\to{1\over2}\left(\begin{array}{cc}\sigma(x) & 0 \\0 & \sigma(x)\end{array}\right)
.\end{align}
It is easy to see that the contribution to the quartic term $\frac{1}{4}\lambda \sigma(x)^4$ is
\begin{align}
\lambda={1\over2}\sum_{i=1}^5(\kappa_i+2\hat{\kappa}_i)
.\label{e14}\end{align}
Therefore $\lambda\to0$ when $\hat{\kappa}_i\to-\frac{1}{2}\kappa_i$ for $i=1..5$, and the mass of this scalar vanishes in this limit for fixed $v$, since $m_\sigma^2\approx2v^2\lambda$. Thus a light scalar emerges if a certain approximate relation exists between those diagrams where the four scalar fields couple to one or two $q'$ loops respectively (corresponding to the one and two trace terms). Meanwhile, as familiar from the 2-Higgs-doublet model, if the $m_3^2$ mass mixing term in (\ref{e2}) is large it gives a large mass to the other physical states. We shall refer to the combination of these two phenomena as the decoupling limit for a light scalar emerging from condensing $t'$ and $b'$ quarks.

We can now start to see the type of dynamics that is required to have a light scalar. From large $N_c$ arguments the two trace terms are ${\cal O}(1/N_c)$ suppressed relative to the one trace terms. In fact in Nambu-Jona-Lasinio models the two trace terms are typically ignored by invoking this large $N_c$ argument. The $\kappa_2,...\kappa_5$ terms are also usually not considered and $\kappa
_1$ is estimated to be large \cite{Hashimoto:2009ty}, so this precludes a light scalar. Similarly there is no light scalar in QCD or QCD-like technicolor theories. The strong interactions must be far away from a large $N_c$ limit to allow for a significant cancellation between the one and two trace terms, and thus we are led to consider a strong $U(1)$ gauge group. (Normal color is an effective flavor with respect to this strong interaction and a two $q'$ loop diagram has a flavor factor of three relative to a one $q'$ loop diagram.) Purely structurally a $U(1)$ may be the only choice for a new gauge interaction that acts on a fourth family, and possibly also the third \cite{Holdom:2011fc}. We take it to be broken near the TeV scale so as to allow the heavy fermions to mix with lighter fermions. The possible fixed point behavior of a strong $U(1)$ at large $N_f$ \cite{Holdom:2010qs,Shrock:2013cca} may also be of interest in this context.

The other ingredient of a decoupling limit is the required $\Phi_{t'}^\dagger\Phi_{b'}$ term. It can be seen that ${\rm Tr}(\tilde{\cal U}^\dagger{\cal U})+{\rm h.c.}$ is a bosonized version of the operator $(i\tau_2)_{ab}(i\tau_2)_{cd}\bar{q}'_{La} q'_{Rc}\bar{q}'_{Lb} q'_{Rd}+{\rm h.c.}$, and so this operator must be present in the underlying theory. Similarly a replacement of a ${\cal U}$ with a $\tilde{\cal U}$ in a quartic term corresponds to an insertion of this operator. This operator cannot be generated perturbatively, and so it represents another distinct feature of the nonperturbative dynamics. This operator may contribute to the breakdown of the $U(1)$, depending on $U(1)$ charge assignments \cite{Holdom:2011fc}. Four-fermion operators of this chirality changing structure have been argued to play a useful role in the generation of other quark and lepton masses \cite{Holdom:1995fu,Holdom:1997hc}, and in particular the top mass \cite{Holdom:1994cr}.

We now turn to the couplings between the $\Phi_{t'}$, $\Phi_{b'}$ fields and the $\Phi_{t}$, $\Phi_{\tau'}$ fields, where these couplings can be treated as $SU(2)_L\times SU(2)_R$ symmetry breaking effects. This will lead to corrections to the decoupling limit that are of order $1/t_\chi$. We can define additional $2\times2$ fields, ${\cal X}_{t}=(\tilde{\Phi}_{t}\;0)$ and ${\cal X}_{\tau'}=(0\;\Phi_{\tau'})$, and thus obtain additional mass mixing terms in the scalar potential,
\begin{align}
&-\lbrace m_4^2{\rm Tr}({\cal U}^\dagger {\cal X}_{t})+m_5^2{\rm Tr}(\tilde{\cal U}^\dagger {\cal X}_{t})+m_6^2{\rm Tr}({\cal X}^\dagger_{\tau'}\tilde {\cal U})+m_7^2{\rm Tr}({\cal X}^\dagger_{\tau'} {\cal U})+{\rm h.c.}\rbrace\nonumber\\
&=-\lbrace m_4^2\Phi_t^\dagger\Phi_{t'}+m_5^2\Phi_t^\dagger\Phi_{b'}+m_6^2\Phi_{\tau'}^\dagger\Phi_{t'}+m_7^2\Phi_{\tau'}^\dagger\Phi_{b'}+{\rm h.c.}\rbrace
.\end{align}
Each of these mass terms again has a corresponding four-fermion interaction in the underlying theory. We can also consider the quartic terms that are linear in the $\Phi_{t}$, $\Phi_{\tau'}$ fields. There are again one and two trace terms.
\begin{align}
&\kappa_6{\rm Tr}({\cal U}^\dagger{\cal U}{\cal U}^\dagger {\cal X}_t)+\kappa_7{\rm Tr}(\tilde{\cal U}^\dagger\tilde{\cal U}\tilde{\cal U}^\dagger{\cal X}_t)+\kappa_8{\rm Tr}({\cal U}^\dagger\tilde{\cal U}{\cal U}^\dagger {\cal X}_t)+\kappa_9{\rm Tr}(\tilde{\cal U}^\dagger{\cal U}\tilde{\cal U}^\dagger {\cal X}_t)\nonumber\\
&\kappa_{10}{\rm Tr}({\cal U}^\dagger{\cal U}\tilde{\cal U}^\dagger {\cal X}_t)+\kappa_{11}{\rm Tr}(\tilde{\cal U}^\dagger{\cal U}{\cal U}^\dagger {\cal X}_t)+\kappa_{12}{\rm Tr}(\tilde{\cal U}^\dagger\tilde{\cal U}{\cal U}^\dagger {\cal X}_t)+\kappa_{13}{\rm Tr}({\cal U}^\dagger\tilde{\cal U}\tilde{\cal U}^\dagger {\cal X}_t)\nonumber\\
&+[\mbox{2 trace terms with }\kappa_i\to\hat\kappa_i]+{\rm h.c.}
\label{e12}\end{align}
Here we have only shown the terms involving ${\cal X}_t$; there are an analogous set of terms involving ${\cal X}_{\tau'}$.

We note that $\hat{\kappa}_i\approx-\frac{1}{2}\kappa_i$ for $i=1,...5$ does not necessarily imply that $\hat{\kappa}_i\approx-\frac{1}{2}\kappa_i$ for $i=6,7,...13$. Terms with a ${\cal X}_t$ have a top loop in addition to the $q'$ loops in the underlying diagrams and if $t$ and $q'$ have opposite $U(1)$ charges \cite{Holdom:2011fc} then it is especially clear that strong $U(1)$ interactions will cause the relative size of the $\kappa_i$ and $\hat{\kappa}_i$ terms to change for $i=6,7,...13$.

We now give the multi-Higgs potential in conventional form, and then we can relate the standard quartic couplings, the $\lambda_i$'s, to the $\kappa_i$'s and $\hat{\kappa}_i$'s. For the quartic terms we only keep terms to first order in the $\Phi_t$, $\Phi_{\tau'}$ fields as these will be sufficient for the leading ${\cal O}(1/t_\chi)$ corrections. (We have included the $m_8^2$ term just to show its effect.) We also ignore charge parity (\textit{CP}) violation and thus assume all coefficients are real.
\begin{align}{\cal V}&=m_0^2\Phi_t^\dagger\Phi_t+m_1^2\Phi_{\tau'}^\dagger\Phi_{\tau'}+m_2^2(\Phi_{t'}^\dagger\Phi_{t'}+\Phi_{b'}^\dagger\Phi_{b'})\nonumber\\
&-\lbrace m_3^2\Phi_{t'}^\dagger\Phi_{b'}+m_4^2\Phi_t^\dagger\Phi_{t'}+m_5^2\Phi_t^\dagger\Phi_{b'}+m_6^2\Phi_{\tau'}^\dagger\Phi_{t'}+m_7^2\Phi_{\tau'}^\dagger\Phi_{b'}+m_8^2\Phi_{\tau'}^\dagger\Phi_{t}+{\rm h.c.}\rbrace\nonumber\\
&+\frac{1}{2}\lambda_2[(\Phi_{t'}^\dagger\Phi_{t'})^2+(\Phi_{b'}^\dagger\Phi_{b'})^2]+\lambda_3(\Phi_{t'}^\dagger\Phi_{t'})(\Phi_{b'}^\dagger\Phi_{b'})+\lambda_4(\Phi_{t'}^\dagger\Phi_{b'})(\Phi_{b'}^\dagger\Phi_{t'})+\lbrace\frac{1}{2}\lambda_5(\Phi_{t'}^\dagger\Phi_{b'})^2+{\rm h.c.}\rbrace\nonumber\\
&+\lbrace[\lambda_6\Phi_{t'}^\dagger\Phi_{b'}+\lambda_7\Phi_t^\dagger\Phi_{t'}+\lambda_8\Phi_t^\dagger\Phi_{b'}+\lambda_9\Phi_{\tau'}^\dagger\Phi_{t'}+\lambda_{10}\Phi_{\tau'}^\dagger\Phi_{b'}](\Phi_{t'}^\dagger\Phi_{t'}+\Phi_{b'}^\dagger\Phi_{b'})\nonumber\\
&+\lambda_{11}\Phi_t^\dagger\Phi_{t'}\Phi_{t'}^\dagger\Phi_{b'}+\lambda_{12}\Phi_t^\dagger\Phi_{b'}\Phi_{b'}^\dagger\Phi_{t'}+\lambda_{13}\Phi_{\tau'}^\dagger\Phi_{t'}\Phi_{t'}^\dagger\Phi_{b'}+\lambda_{14}\Phi_{\tau'}^\dagger\Phi_{b'}\Phi_{b'}^\dagger\Phi_{t'}\nonumber\\
&+\lambda_{15}\Phi_t^\dagger\Phi_{t'}\Phi_{b'}^\dagger\Phi_{t'}+\lambda_{16}\Phi_t^\dagger\Phi_{b'}\Phi_{t'}^\dagger\Phi_{b'}+\lambda_{17}\Phi_{\tau'}^\dagger\Phi_{t'}\Phi_{b'}^\dagger\Phi_{t'}+\lambda_{18}\Phi_{\tau'}^\dagger\Phi_{b'}\Phi_{t'}^\dagger\Phi_{b'}+{\rm h.c.}\rbrace+...
\end{align}
We obtain the following relations for the terms that only involve $\Phi_{t'}$ and $\Phi_{b'}$ fields.
\begin{align}
&\lambda_2=\lambda_3=2\kappa_1+2\hat{\kappa}_1+2\hat{\kappa}_{2}\nonumber\\
&\lambda_4=-2\kappa_1+2\kappa_{2}+2\kappa_3+4\hat{\kappa}_3\nonumber\\
&\lambda_5=2\kappa_{5}+4\hat{\kappa}_{5}\nonumber\\
&\lambda_6={1\over2}\kappa_{4}+\hat{\kappa}_{4}
\label{e3}\end{align}
 In the decoupling limit where $\hat{\kappa}_i\approx-\frac{1}{2}\kappa_i$ we have $\lambda_2=\lambda_3\approx-\frac{1}{2}\lambda_4\approx(\kappa_1-\kappa_2)$ and $\lambda_5\approx\lambda_6\approx0$. It might be expected that both $\kappa_1$ and $\kappa_2$ are positive and that the two insertions of $\tilde{\cal U}$ in the $\kappa_2$ term result in $\kappa_1-\kappa_2>0$. For the quartic terms involving $\Phi_t$ we have
\begin{align}
&\lambda_7=\kappa_6+\hat{\kappa}_6+\hat\kappa_{12},\quad\lambda_8=\kappa_7+\hat{\kappa}_7+\hat\kappa_{10},\nonumber\\
&\lambda_{11}=-\kappa_7+\kappa_{10}+\kappa_{11}+2\hat\kappa_{11},\quad\lambda_{12}=-\kappa_6+\kappa_{12}+\kappa_{13}+2\hat\kappa_{13},\nonumber\\
&\lambda_{15}=\kappa_{8}+2\hat\kappa_{8},\quad\lambda_{16}=\kappa_{9}+2\hat\kappa_{9}.
\label{e22}\end{align}
There are analogous relations for the quartic terms involving $\Phi_{\tau'}$.

We use the minimization conditions to eliminate to $m_0$, $m_1$ and $m_2$ and thus write the mass matrices,
\begin{align}
{\cal M}^2_\eta&=M+E\\
{\cal M}^2_{h^\pm}&={\cal M}^2_\eta+C\\
{\cal M}^2_h&={\cal M}^2_\eta+S
,\end{align}
as follows, with [$t$, $t'$, $b'$, $\tau'$] as the basis order and where it is understood that the matrices are symmetric.
\begin{align}
M=\left(\begin{array}{cccc}
t_\chi(m_4^2+m_5^2)+m_8^2 & -m_4^2 & -m_5^2 & -m_8^2 \\
  & m_3^2+\frac{m_4^2+m_6^2}{t_\chi} & -m_3^2 & -m_6^2 \\
 &  & m_3^2+\frac{m_5^2+m_7^2}{t_\chi} & -m_7^2\\
 &  &  & t_\chi(m_6^2+m_7^2)+m_8^2
\end{array}\right)
\label{e6}\end{align}
\begin{align}
E&=v'^2\left(\begin{array}{cccc}
-t_\chi\lambda_a & \lambda_7+{1\over2}\lambda_{12}+\lambda_{15}-{1\over2}\lambda_{16} & \lambda_8+{1\over2}\lambda_{11}+\lambda_{16}-{1\over2}\lambda_{15} & 0 \\
 & -\lambda_5-\lambda_6-{\lambda_c\over t_\chi} & \lambda_5+\lambda_6+{\lambda_g\over t_\chi} & \lambda_9+{1\over2}\lambda_{14}+\lambda_{17}-{1\over2}\lambda_{18}   \\
 &  &-\lambda_5-\lambda_6-{\lambda_d\over t_\chi} & \lambda_{10}+{1\over2}\lambda_{13}+\lambda_{18}-{1\over2}\lambda_{17}  \\
 & &  & -t_\chi\lambda_b
\end{array}\right)
\label{e7}\end{align}
\begin{align}
C=\frac{v'^2}{2}\left(\begin{array}{cccc}
0 & \lambda_{11}-\lambda_{12}-\lambda_{15}+\lambda_{16} & \lambda_{12}-\lambda_{11}+\lambda_{15}-\lambda_{16} & 0 \\
 & -\lambda_4+\lambda_5-2\frac{\lambda_{11}+\lambda_{13}-\lambda_{15}-\lambda_{17}}{t_\chi} & \lambda_4-\lambda_5+{\lambda_f-\lambda_g\over t_\chi} & \lambda_{13}-\lambda_{14}-\lambda_{17}+\lambda_{18}  \\
 &  & -\lambda_4+\lambda_5-2\frac{\lambda_{12}+\lambda_{14}-\lambda_{16}-\lambda_{18}}{t_\chi}& \lambda_{14}-\lambda_{13}+\lambda_{17}-\lambda_{18}  \\
 & & & 0
\end{array}\right)
\end{align}
\begin{align}
S=v'^2&\left(\begin{array}{cccc}
0 & \lambda_7+\lambda_8+\lambda_{11}+\lambda_{16} & \lambda_7+\lambda_8+\lambda_{12}+\lambda_{15} & 0 \\
 & \lambda_2+\lambda_5+2\lambda_6+2\frac{\lambda_7+\lambda_9+\lambda_{15}+\lambda_{17}}{t_\chi} & \lambda_3+\lambda_4+2\lambda_6+{\lambda_e+\lambda_f\over t_\chi}  & \lambda_9+\lambda_{10}+\lambda_{13}+\lambda_{18} \\
 &  & \lambda_2+\lambda_5+2\lambda_6+2\frac{\lambda_8+\lambda_{10}+\lambda_{16}+\lambda_{18}}{t_\chi} & \lambda_9+\lambda_{10}+\lambda_{14}+\lambda_{17} \\
 & & & 0
\end{array}\right)
\label{e8}\end{align}
We have defined
\begin{align}
&\lambda_a=\lambda_7+\lambda_8+{1\over2}(\lambda_{11}+\lambda_{12}+\lambda_{15}+\lambda_{16}),\quad\lambda_b=\lambda_9+\lambda_{10}+{1\over2}(\lambda_{13}+\lambda_{14}+\lambda_{17}+\lambda_{18}),\nonumber\\
&\lambda_c=\lambda_7+\lambda_{9}+{1\over2}(\lambda_{12}+\lambda_{14}+4\lambda_{15}+\lambda_{16}+4\lambda_{17}+\lambda_{18}),\nonumber\\
&\lambda_d=\lambda_{8}+\lambda_{10}+{1\over2}(\lambda_{11}+\lambda_{13}+\lambda_{15}+4\lambda_{16}+\lambda_{17}+4\lambda_{18})\nonumber,\\
&\lambda_e=\lambda_7+\lambda_8+\lambda_9+\lambda_{10},\quad\lambda_f=\lambda_{11}+\lambda_{12}+\lambda_{13}+\lambda_{14},\quad\lambda_g=\lambda_{15}+\lambda_{16}+\lambda_{17}+\lambda_{18}
.\label{e17}\end{align}

Let us start by turning off all the terms in $\cal V$ that are linear in the $\Phi_t$, $\Phi_{\tau'}$ fields and only consider the $2\times2$ version of the above matrices for the $\Phi_{t'}$, $\Phi_{b'}$ sector. These results will then be familiar from the 2-Higgs-doublet model with $\tan\beta=1$. The state $h_1=(\sigma_{t'}+\sigma_{b'})/\sqrt{2}$ has mass
\begin{align}
m_{h_1}^2&=v'^2(\lambda_2+\lambda_3+\lambda_4+\lambda_5+4\lambda_6)\nonumber\\
&=2v'^2\sum_{i=1}^5(\kappa_i+2\hat{\kappa}_i)
\label{e5}\end{align}
where we have used the relations in (\ref{e3}). A neutral pseudoscalar, $\eta_1=(\eta_{t'}-\eta_{b'})/\sqrt{2}$,\footnote{This combination is a isosinglet due to the definition of the fields in (\ref{e4}).} has mass
\begin{align}
m_{\eta_1}^2=2m^2_3-v'^2(2\lambda_5+2\lambda_6)
,\end{align}
a charged scalar $h^\pm=(\phi^\pm_{t'}-\phi^\pm_{b'})/\sqrt{2}$ has mass
\begin{align}
m_{h^\pm}^2=2m^2_3-v'^2(\lambda_4+\lambda_5+2\lambda_6)
,\end{align}
and the next heavier neutral scalar $h_2=(\sigma_{t'}-\sigma_{b'})/\sqrt{2}$ has mass
\begin{align}
m_{h_2}^2=2m^2_3+v'^2(\lambda_2-\lambda_3-\lambda_4-\lambda_5-2\lambda_6)
.\end{align}
$SU(2)_L\times SU(2)_R$ symmetry implies $\lambda_2=\lambda_3$ and thus a near degeneracy in the $h_2$ and $h^\pm$ masses. In the decoupling limit
\begin{align}
m_{h^\pm}^2\approx m_{h_2}^2 \approx m_{\eta_1}^2+2v'^2(\kappa_1-\kappa_2)
.\label{e15}\end{align}

We also give the trilinear couplings involving $h_1$ that occur at leading order in $1/t_\chi$.
\begin{align}
\frac{v'}{2\sqrt{2}}&\left[\frac{m_{h_1}^2}{v'^2}h_1^3+(3\lambda_2-\lambda_3-\lambda_4-\lambda_5)h_1h_2^2\right.\nonumber
\\&\left.\mbox{  }+2(\lambda_2+\lambda_3-\lambda_4-\lambda_5)h_1h^+h^-+(\lambda_2+\lambda_3+\lambda_4-3\lambda_5)h_1\eta_1^2\right]
\end{align}
The $h_1^3$ coupling is the SM value and the other couplings in the decoupling limit reduce to
\begin{align}
\approx\sqrt{2}v'(\kappa_1-\kappa_2)[h_1(h_2^2+2h^+h^-)]
.\end{align}
From (\ref{e15}) this coupling is related to a mass difference. The $h_1 h^+h^-$ couplings imply a charged scalar loop correction to $h_1\to\gamma\gamma$, but this correction is further suppressed by $v'^2/m_{h^\pm}^2$.

Now let us turn back on the mixing between the $\Phi_{t'}$, $\Phi_{b'}$ and the $\Phi_t$, $\Phi_{\tau'}$ sectors and so study the ${\cal O}(1/t_\chi)$ corrections. The full matrices ${\cal M}^2_\eta$ and ${\cal M}^2_{h^\pm}$ each have a vanishing eigenvalue (the Goldstone mode) with the same eigenvector, $\frac{\sin\chi}{\sqrt{2}}[1/t_\chi,1,1,1/t_\chi]$. These matrices have the same form as $M$, that is they can be written as a matrix, $M$, with the masses redefined to include the $\lambda_i$ contributions. Another eigenvalue of $M$ is
\begin{align}
&2m_3^2+\frac{2}{t_\chi}\left[\frac{m_4^2m_5^2}{(m_4^2+m_5^2)}+\frac{m_6^2m_7^2}{(m_6^2+m_7^2)}\right]+{\cal O}(\frac{1}{t_\chi^2})
.\label{e18}\end{align}
This is the next lowest eigenvalue if it is less than $t_\chi(m_4^2+m_5^2)$ and $t_\chi(m_6^2+m_7^2)$, which are respectively the other two eigenvalues at leading order. The ${\cal O}(1/t_\chi)$ corrections for $m_{\eta_1}^2$ and $m_{h^\pm}^2$ can be obtained by using (\ref{e18}) with masses suitably redefined to represent the matrices ${\cal M}^2_\eta$ and ${\cal M}^2_{h^\pm}$.

Of more interest is the neutral scalar mass matrix ${\cal M}_h^2$ that is obtained from ${\cal M}^2_\eta$ by adding $S$; this raises the vanishing eigenvalue and distorts the corresponding eigenvector. We have already seen how this eigenvalue can remain small compared to the next higher eigenvalues of all three mass matrices.  We now need to see how the corresponding eigenvector can be close to $\frac{\sin\chi}{\sqrt{2}}[-1/t_\chi,1,1,1/t_\chi]$. First we see that the existence of such an eigenvector constrains the relevant mixing terms in the mass matrix,
\begin{align}
\frac{{\cal M}^2_{h12}+{\cal M}^2_{h13}}{{\cal M}^2_{h11}}\approx \frac{1}{t_\chi},\quad\quad
\frac{{\cal M}^2_{h24}+{\cal M}^2_{h34}}{{\cal M}^2_{h44}}\approx -\frac{1}{t_\chi}.
\label{e16}\end{align}
By inspection of the mass matrix this then leads to the necessary constraints,
\begin{align}
\lambda_a v'^2\approx (m_4^2+m_5^2)/2\mbox{ and }\lambda_b v'^2\approx 0
,\end{align}
where $\lambda_a$ and $\lambda_b$ are defined in (\ref{e17}). We note that $\lambda_a v'^2$ and $\lambda_b v'^2$ cannot be greater than $m_4^2+m_5^2$ and $m_6^2+m_7^2$, respectively, to ensure that ${\cal M}^2_{h11}$ and ${\cal M}^2_{h44}$ are positive.

With this we can obtain the ${\cal O}(1/t_\chi)$ correction to $m^2_{h_1}$. There are contributions both from the mixing as described by (\ref{e16}) (this reduces $m^2_{h_1}$) and from the $1/t_\chi$ corrections that are present in the inner $2\times2$ block. The combined correction is found to give the $1/t_\chi$ term in the following.
\begin{align}
m_{h_1}^2&=v'^2(\lambda_2+\lambda_3+\lambda_4+\lambda_5+4\lambda_6)+\frac{1}{t_\chi}(m_4^2+m_5^2)+{\cal O}(\frac{1}{t_\chi^2})\label{e21}\nonumber\\
&=2v'^2\sum_{i=1}^5(\kappa_i+2\hat{\kappa}_i)+\frac{v'^2}{t_\chi}\sum_{i=6}^{13}(\kappa_i+2\hat{\kappa}_i)+{\cal O}(\frac{1}{t_\chi^2})
\end{align}
The second sum is a representation of $2\lambda_a$ after using (\ref{e22}). For illustration if we take $t_\chi(m_3^2+m_4^2)=1$ TeV and $t_\chi=5$ then the second term is (200 GeV)$^2$. For a 126 GeV mass Higgs boson the first term would need to be $-$(155 GeV)$^2$, or in other words the sum of $\lambda_i$s in the first term is $-0.8$. In this case the $1/t_\chi$ corrections are stabilizing the vacuum.

We may also obtain the $1/t_\chi$ correction to the $h_1^3$ coupling,
\begin{align}
\frac{v'}{2\sqrt{2}}h_1^3\left[\lambda_2+\lambda_3+\lambda_4+\lambda_5+4\lambda_6+\frac{1}{t_\chi}\lambda_h+{\cal O}(\frac{1}{t_\chi^2})\right]
,\end{align}
with $\lambda_h=2\lambda_7+2\lambda_8-4\lambda_9-4\lambda_{10}+\lambda_{11}+\lambda_{12}-2\lambda_{13}-2\lambda_{14}+\lambda_{15}+\lambda_{16}-2\lambda_{17}-2\lambda_{18}$. This $1/t_\chi$ correction may also be relatively significant when compared to the anomalously small lowest order value.

We would now like to explore just how close we need to be to the mixing pattern we have described, that is how close to the relations in (\ref{e16}) we need to be, for consistency with the present data. For this it is sufficient to simplify things and set ${\cal M}^2_{h11}={\cal M}^2_{h44}\equiv\mu^2$ and ${\cal M}^2_{h14}={\cal M}^2_{h41}=0$ [notice how (\ref{e18}) does not depend on $m_8^2$]. The difference between ${\cal M}^2_{h22}$ and ${\cal M}^2_{h33}$ is suppressed by $1/t_\chi$ and it could be of either sign, so we also set ${\cal M}^2_{h22}={\cal M}^2_{h33}$.  We can then write
\begin{align}
{\cal M}^2_h=\mu^2\left(\begin{array}{cccc}1 & -d & -c & 0 \\-d & a & -b & -f \\-c & -b & a & -e\\0 & -f & -e & 1\end{array}\right)
.\label{e13}\end{align}
Insisting that all eigenvalues are positive implies that
\begin{align}
&a>A\equiv\frac{1}{2}(c^2+d^2+e^2+f^2),\\
&-(a-A)\lesssim b+cd+ef\lesssim a-A
.\label{e19}\end{align}
The decoupling limit corresponds to when $b$ is near the upper end of its range in (\ref{e19}). The relations in  (\ref{e16}) become
\begin{align}
c+d\approx -1/t_\chi,\quad e+f\approx 1/t_\chi.
\label{e20}\end{align}

The two lowest eigenvalues (as long as $a+b\lesssim1$) are approximately
\begin{align}
&{m}_{h_1}^2/\mu^2\approx a-b-\frac{(c+d)^2+(e+f)^2}{2}
,\label{e9}\\
&{m}_{h_2}^2/\mu^2\approx a+b-\frac{(c-d)^2+(e-f)^2}{2}
.\end{align}
The corresponding eigenvectors are approximately proportional to
\begin{align}
&[c+d,1-\frac{(c^2+e^2-d^2-f^2)}{4b},1+\frac{(c^2+e^2-d^2-f^2)}{4b},e+f],
\label{e10}\\
&[\frac{d-c}{1-2b},1+\frac{(c^2+e^2-d^2-f^2)}{4b(1-2b)},-1+\frac{(c^2+e^2-d^2-f^2)}{4b(1-2b)},\frac{f-e}{1-2b}]
.\label{e11}\end{align}
The new correction terms in (\ref{e10}) will affect the $h_1$ coupling to $\gamma\gamma$, but will cancel in the $gg$ and $VV$ ($WW$ and $ZZ$) couplings. The couplings of $h_2$ are determined approximately by (\ref{e11}) and in particular we see that its $VV$ couplings are quite suppressed. The $h_2$ coupling to $t\bar t$ or $\tau'\tau'$ depends on $d-c$ or $f-e$ respectively and the $gg$ coupling is also strongly dependent on these differences. In addition the coupling responsible for the decay $h_2\to h_1 h_1$ only appears at order $1/t_\chi$, and so $h_2\to\eta_1 Z$ may be a dominant decay of $h_2$ if $m_{h_2}-m_{\eta_1}$ is large enough.

To find the allowed region in the space of $c,d,e,f$ parameters we perform a scan over this space for the fixed $a$ and $b$. Rather than use the approximate results in (\ref{e9})-(\ref{e11}) we instead use the eigenvalues and eigenvectors of (\ref{e13}) as obtained numerically. We uniformly sample $c,d,e,f$ and only keep those values that produce positive mass squares and are such that the $h_1$ production cross section times width ($\sigma\times\Gamma$) into $VV$, $\gamma\gamma$ and $\tau\tau$ respectively is each within $20\%$ of the SM Higgs value (the $h_1$ coupling to $\tau\tau$ is assumed to be $s_{\tau'}v/v_{\tau'}$ times the SM value). Then the acceptable values of $c,d,e,f$ are shown in Fig.~1, where we have made the choice $a=0.4$, $b=0.35$ and $t_\chi=5$. This shows quite clearly the extent to which the sums $c+d$ and $e+f$ are constrained, while showing that the differences $c-d$ and $e-f$ are not constrained. The resulting values for $\sigma\times\Gamma$ for $VV$, $\gamma\gamma$ and $\tau\tau$ are spread quite uniformly over the allowed ranges.
\begin{figure}[t]
\vspace{-0ex}
\centering\includegraphics[scale=.3]{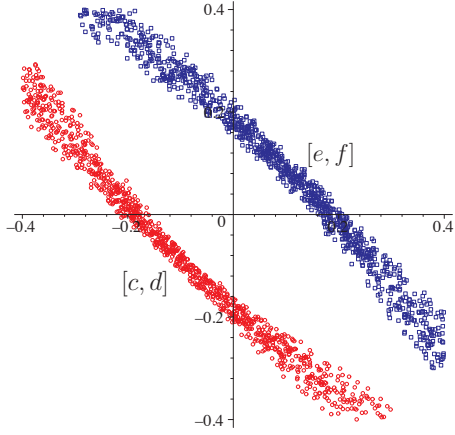}
\vspace{-0ex}
\caption{The allowed ranges of the parameters $c,d,e,f$ appearing in the scalar mass matrix ${\cal M}^2_h$ in (\ref{e13}) for $a=0.4$, $b=0.35$ and $t_\chi=5$.}\end{figure}
\begin{figure}[h]
\vspace{-0ex}
\centering\includegraphics[scale=.29]{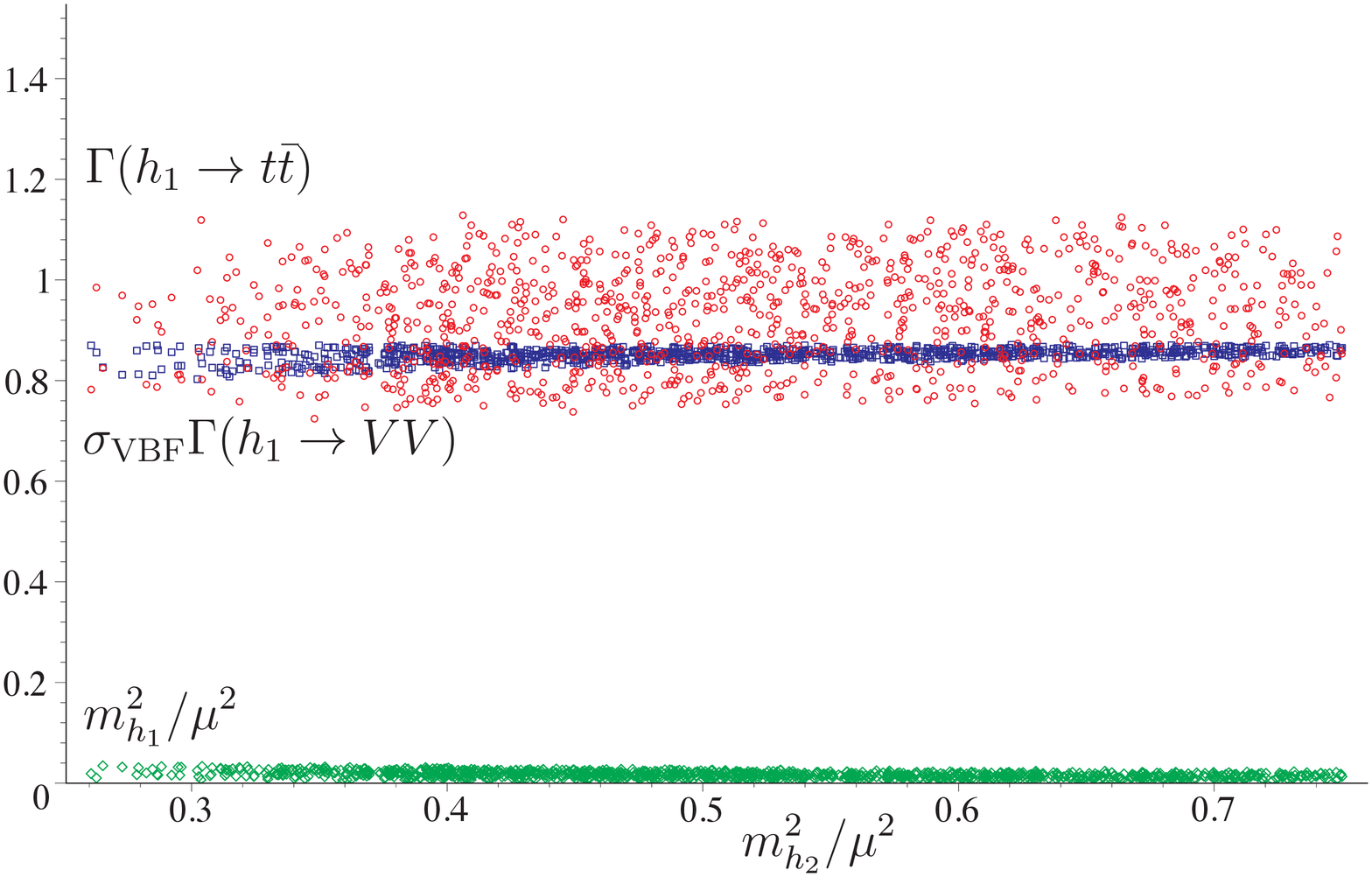}
\centering\includegraphics[scale=.29]{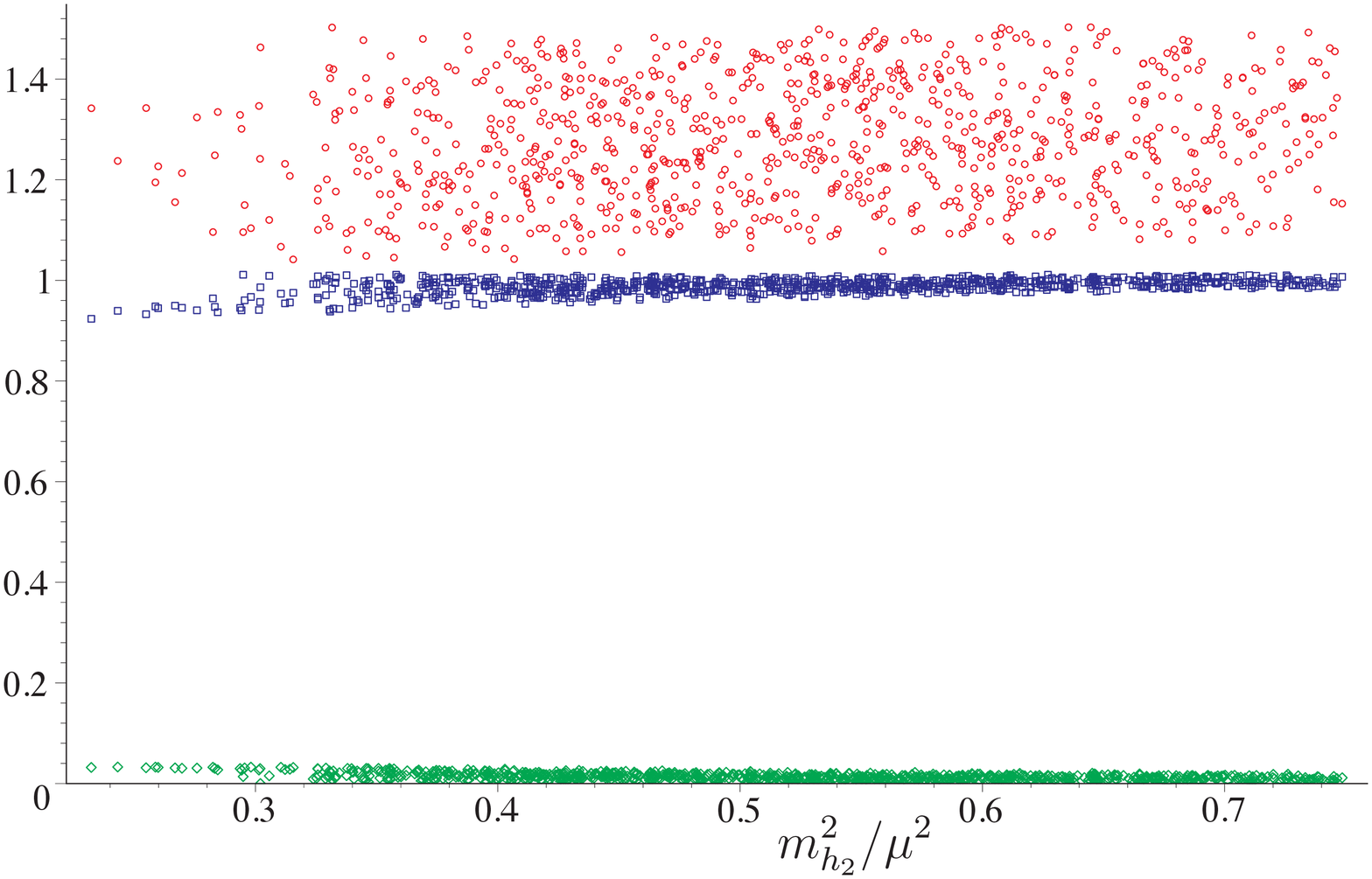}
\vspace{-5ex}
\caption{$\sigma_{VBF}\Gamma(h_1\to VV)$ (blue squares) and $g^2_{h_1\bar tt}$ (red circles), both relative to the SM Higgs boson, and $m_{h_1}^2/\mu^2$ (green diamonds), as a function of $m_{h_2}^2/\mu^2$. The left and right plots are for two different scans as described in the text.}\end{figure}

In Fig.~2a we display some quantities as a function of $m_{h_2}^2/\mu^2$, where the large range of $m_{h_2}^2/\mu^2$ is due to the variation in $c-d$ and $e-f$. If these differences were small then $m_{h_2}^2/\mu^2$ would be confined to the upper end of its range. The figure shows that the values of $m_{h_1}^2/\mu^2$ are small (with an average value of $0.016$) and are quite independent of the allowed values of $c,d,e,f$, as (\ref{e20}) and (\ref{e9}) indicate. This figure also displays (1) $\sigma\times\Gamma$ for the vector boson fusion (VBF) or associated production (\textit{VH}) process with $h_1$ decay to $VV$ and (2) the square of the $h_1$ coupling to $t\bar t$,  both relative to the SM Higgs boson. The former is seen to have values that are about 0.85 times the SM values; this is a reflection of the slightly smaller $h_1VV$ coupling we mentioned earlier. This result can receive corrections from possible dimension-six terms in the effective scalar doublet theory.

As another possibility we point out a significant leeway that is still permitted by the data, which allows various Higgs couplings to be uniformly smaller (or larger) than in the SM. We show a fit to the combined data using HiggsSignals 1.2 \cite{Bechtle:2013xfa} in Fig.~3. It displays the correlation in the allowed scaling of the $b\bar{b}$ coupling (not squared) with the allowed uniform scaling of the $\gamma\gamma$, $\tau\tau$ and $VV$ couplings. The reason for this is that the Higgs widths into each of $\gamma\gamma$, $\tau\tau$ and $VV$ can all be smaller than in the standard model as long as the width into $b\bar{b}$ is also appropriately smaller. The latter reduces the total width and thus boosts the branching ratios up to the observed values. We mentioned earlier that the $h_1 \bar{b}b$ coupling can receive a ${\cal O}(1/t_\chi)$ correction.
\begin{figure}[t]
\vspace{-5ex}
\centering\includegraphics[scale=1]{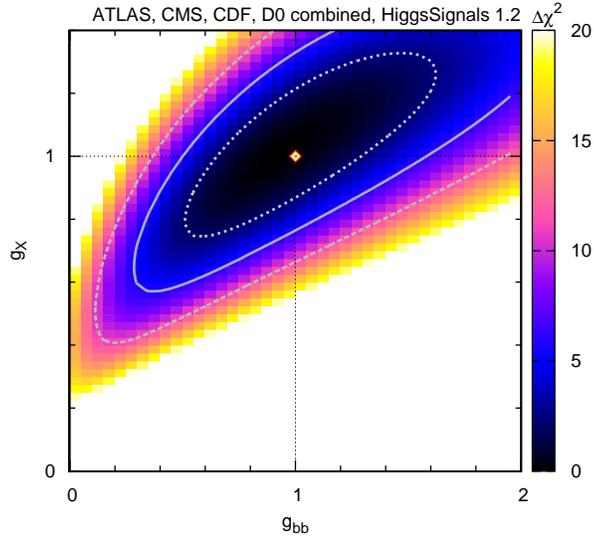}
\vspace{-6ex}
\caption{$g_X$ is a uniform scaling of the Higgs couplings to $VV$, $\gamma\gamma$ and $\tau\tau$ and $g_{bb}$ scales the Higgs coupling to $b\bar b$. [based on data included in HiggsSignals 1.2 (March 2014)].}\end{figure}

The present data can easily accommodate a reduction in $\sigma\times\Gamma$ for $\gamma\gamma$, $\tau\tau$ and $VV$ on the order of $0.85$ that is compensated by a reduced total width due to a smaller $b\bar{b}$ coupling. We thus perform a second scan over the $c,d,e,f$ parameters where we assume that the total width is reduced by $0.85$. The result is shown in Fig.~2b. The VBF or \textit{VH} processes are now SM-like in size, but the square of the $h_1$ coupling to $t\bar t$ is now seen to be enhanced. The average value of $m_{h_1}^2/\mu^2$ is little changed at 0.014.

The couplings of $h_2$ are also determined in the scan and in Fig.~4 we show results for the first scan above (the second scan is similar). In Fig.~4a we show the gluon fusion production cross section for $h_2$. The range of values grows very dramatically as $m_{h_2}^2$ decreases. In Fig.~4b we show $\hat{\Gamma}(h_2\to t\bar t)+\hat{\Gamma}(h_2\to\tau'\tau')$ where $\hat{\Gamma}$ denotes the width relative to the SM Higgs boson. Thus at least one of these widths also grows dramatically as $m_{h_2}^2$ decreases. From this it would appear that a $h_2$ mass that is well below its maximum value could easily be ruled out. Meanwhile $\hat{\Gamma}(h_2\to VV)/(\hat{\Gamma}(h_2\to\bar{t}t)+\hat{\Gamma}(h_2\to\tau'\tau'))$ remains small, remaining below $\approx0.004$ for any $m_{h_2}^2$.
\begin{figure}[t]
\vspace{-0ex}
\centering\includegraphics[scale=.29]{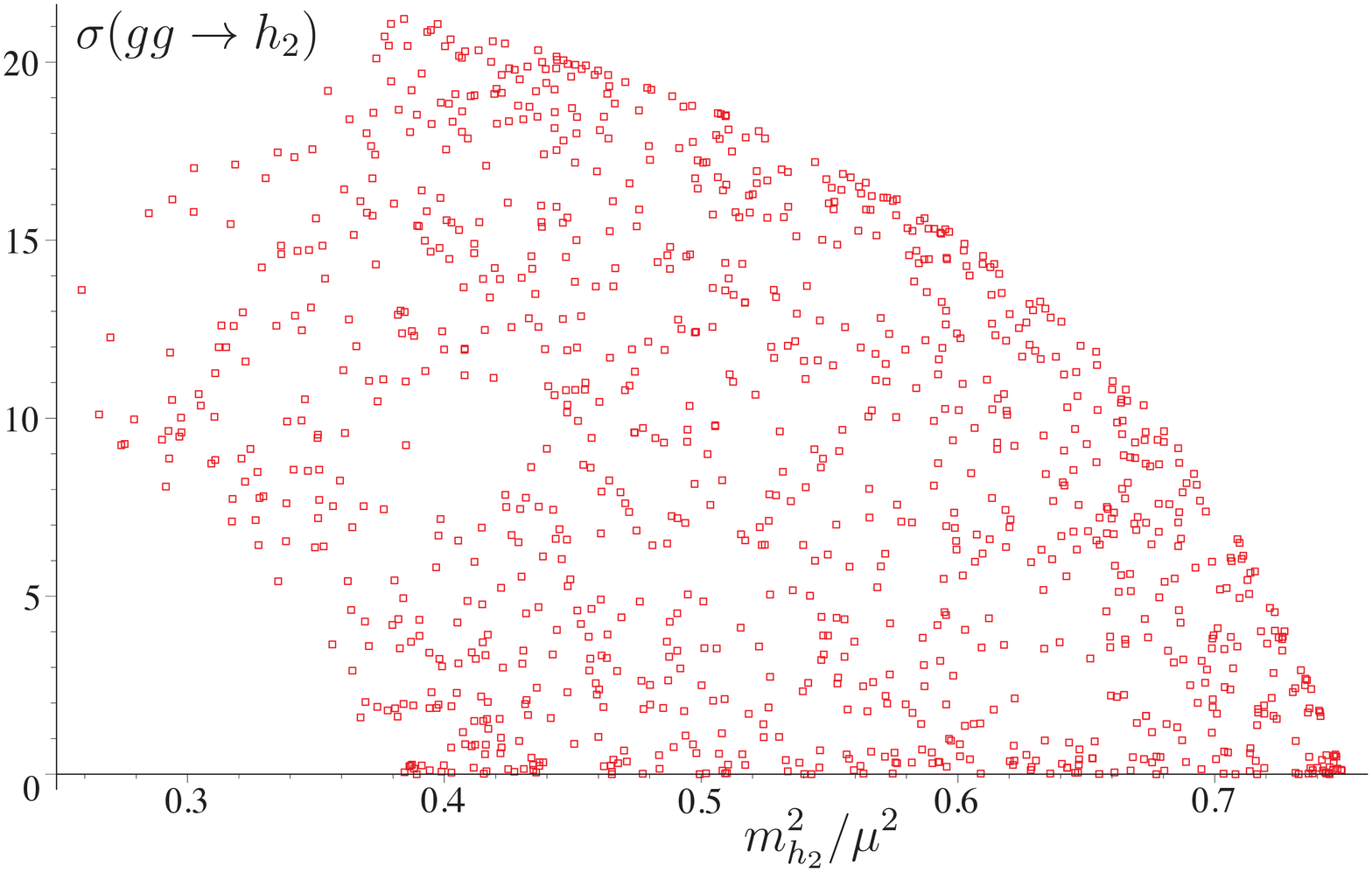}
\centering\includegraphics[scale=.29]{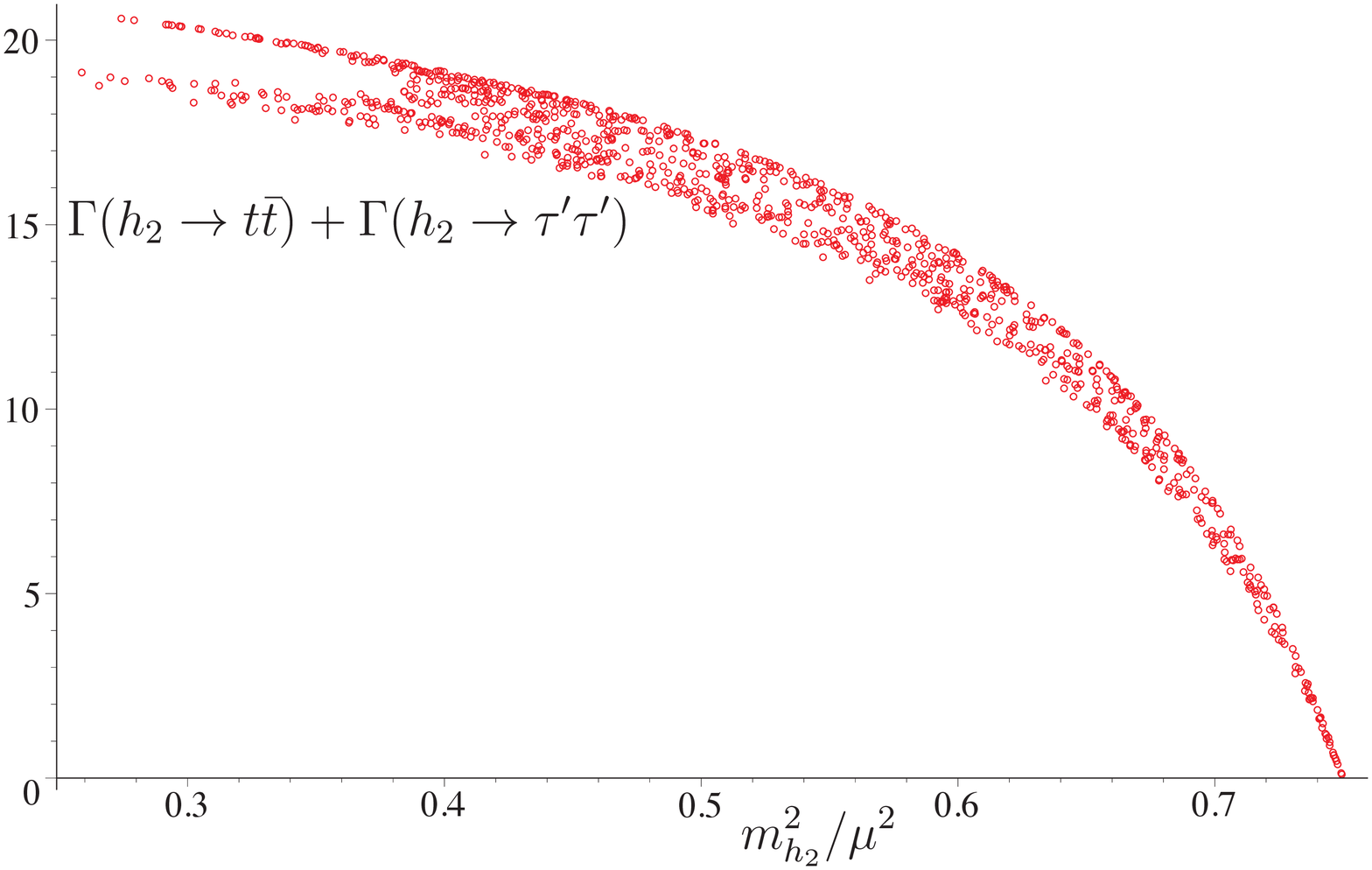}
\vspace{-2ex}
\caption{(a) $\sigma(gg\to h_2)$ and (b) $\Gamma(h_2\to t\bar t)+\Gamma(h_2\to\tau'\tau')$, both relative to the SM Higgs boson.}\end{figure}

The heaviest two neutral scalars $h_3$ and $h_4$ have masses of order $\mu$ (this is $\mu\approx1$ TeV if our illustrative value of $m_{h_1}^2/\mu^2$ is to produce the correct $m_{h_1}$). Their eigenvectors are dominated by the $t$ and $\tau'$ components and so the sum of the squares of the $h_3$ couplings to $t$ and $\tau'$ will be a factor of $\approx 2t_\chi^2$ larger than the square of the SM Higgs coupling to $t$  (and the same for $h_4$). Among the scalars $h_2$, $h_3$ $h_4$, the most interesting one may be the one with the largest product of production cross section and branching ratio to $\tau'\tau'$. With an enhanced cross section and significant branching ratio, such a boson could be accessible even with a large mass.

We have mentioned the Yukawa couplings of the scalar fields to the lighter families and the suppression of the flavor changing couplings of $h_1$ by ${\cal O}(1/t_\chi)$. The heavier scalars can have flavor changing couplings that are not suppressed in this way, although in this case the higher mass of these states can produce a similar suppression. In both cases the ultimate size of these effects will be governed by the form of the Yukawa couplings. These couplings are induced by four-fermion interactions that couple light to heavy fermions and that reflect new flavor physics at scales up to $\approx10^3$ TeV. The structure of the underlying flavor physics may be such as to give rise to additional suppression of flavor changing neutral currents. This can occur through approximate symmetries in the effective theory, of the standard discrete type or of the continuous type \cite{Hadeed:1985xn,Branco:1996bq,Botella:2014ska,Bhattacharyya:2014nja}. Approximate symmetries are consistent because of the natural UV cutoff of loop effects involving scalars.

In summary we have discussed some particular features of strong interactions involving a fourth family that could underlie the existence of a light Higgs-like scalar. We argued that ``small $N_c$'' dynamics is necessary for a partial cancellation between the one and two trace contributions to the lightest scalar mass. This points to a strong and broken $U(1)$ gauge interaction.  The other required feature is a scalar mass mixing term, well known in 2-Higgs-doublet models, that pushes the other states to higher mass. The origin of this term lies with a four-fermion interaction of a certain chiral structure that cannot be generated perturbatively. These features of the strong interactions can allow one to be ``accidentally'' close to a decoupling limit for a light scalar. We also commented on how it can be that a linear sigma model provides a good description of such a scalar.

The small mass of the $t$ and $\tau'$ relative to the large mass of $t'$ and $b'$ implies a similar ratio of the vevs, and this small ratio determines the size of corrections to the decoupling limit. It also implies that the light scalar has small $\sigma_t$ and $\sigma_{\tau'}$ components.  When the relative sign of the $\sigma_t$ component is negative this changes the sign of the scalar coupling to the top quark. We have shown how this can emerge via a 4-Higgs-doublet potential and how it is needed to bring $gg$ and $\gamma\gamma$ couplings into line with the observed values. Experimentally the sign of the Higgs coupling to the top is accessible through the study of Higgs boson plus single top production \cite{Barger:2009ky,Farina:2012xp,Biswas:2013xva,Englert:2014pja,Chang:2014rfa}.

\section*{Acknowledgments}
This work was supported in part by the Natural Science and Engineering Research Council of Canada.


\begin{thebibliography}{99}
\bibitem{Holdom:2006mr} 
  B.~Holdom,
  JHEP {\bf 0608}, 076 (2006)
  [hep-ph/0606146].
\bibitem{Kribs:2007nz} 
  G.~D.~Kribs, T.~Plehn, M.~Spannowsky and T.~M.~P.~Tait,
  Phys.\ Rev.\ D {\bf 76}, 075016 (2007)
  [arXiv:0706.3718 [hep-ph]].
\bibitem{cat} C.~Gomez-Sanchez and B.~Holdom, work in progress.
\bibitem{Belyaev:2013ida} 
  A.~Belyaev, M.~S.~Brown, R.~Foadi and M.~T.~Frandsen,
  arXiv:1309.2097 [hep-ph].
  \bibitem{GarciaMartin:2011jx} 
  R.~Garcia-Martin, R.~Kaminski, J.~R.~Pelaez and J.~Ruiz de Elvira,
  Phys.\ Rev.\ Lett.\  {\bf 107}, 072001 (2011)
  [arXiv:1107.1635 [hep-ph]].
\bibitem{Holdom:1988yj} 
  B.~Holdom,
  Phys.\ Rev.\ Lett.\  {\bf 60}, 1233 (1988).
\bibitem{Hashimoto:2009ty} 
  M.~Hashimoto and V.~A.~Miransky,
  Phys.\ Rev.\ D {\bf 81}, 055014 (2010)
  [arXiv:0912.4453 [hep-ph]].
\bibitem{Holdom:2011fc} 
  B.~Holdom,
  Phys.\ Lett.\ B {\bf 703}, 576 (2011)
  [arXiv:1107.3167 [hep-ph]].
 \bibitem{Holdom:2010qs} 
  B.~Holdom,
  Phys.\ Lett.\ B {\bf 694}, 74 (2010)
  [arXiv:1006.2119 [hep-ph]].
\bibitem{Shrock:2013cca} 
  R.~Shrock,
  Phys.\ Rev.\ D {\bf 89}, 045019 (2014)
  [arXiv:1311.5268 [hep-th]].
\bibitem{Holdom:1995fu} 
  B.~Holdom,
  Phys.\ Rev.\ D {\bf 54}, 1068 (1996)
  [hep-ph/9512298].
\bibitem{Holdom:1997hc} 
  B.~Holdom,
  Phys.\ Rev.\ D {\bf 57}, 357 (1998)
  [hep-ph/9705231].
\bibitem{Holdom:1994cr} 
  B.~Holdom,
  Phys.\ Lett.\ B {\bf 336}, 85 (1994)
  [hep-ph/9407244].
\bibitem{Bechtle:2013xfa} 
  P.~Bechtle, S.~Heinemeyer, O.~Stl, T.~Stefaniak and G.~Weiglein,
  Eur.\ Phys.\ J.\ C {\bf 74}, 2711 (2014)
  [arXiv:1305.1933 [hep-ph]].
\bibitem{Hadeed:1985xn} 
  A.~M.~Hadeed and B.~Holdom,
  Phys.\ Lett.\ B {\bf 159}, 379 (1985).
\bibitem{Branco:1996bq} 
  G.~C.~Branco, W.~Grimus and L.~Lavoura,
  Phys.\ Lett.\ B {\bf 380}, 119 (1996)
  [hep-ph/9601383].
\bibitem{Botella:2014ska} 
  F.~J.~Botella, G.~C.~Branco, A.~Carmona, M.~Nebot, L.~Pedro and M.~N.~Rebelo,
  arXiv:1401.6147 [hep-ph].
\bibitem{Bhattacharyya:2014nja} 
  G.~Bhattacharyya, D.~Das and A.~Kundu,
  Phys.\ Rev.\ D {\bf 89}, 095029 (2014)
  [arXiv:1402.0364 [hep-ph]].
\bibitem{Barger:2009ky} 
  V.~Barger, M.~McCaskey and G.~Shaughnessy,
  Phys.\ Rev.\ D {\bf 81}, 034020 (2010)
  [arXiv:0911.1556 [hep-ph]].
\bibitem{Farina:2012xp} 
  M.~Farina, C.~Grojean, F.~Maltoni, E.~Salvioni and A.~Thamm,
  JHEP {\bf 1305}, 022 (2013)
  [arXiv:1211.3736 [hep-ph]].
\bibitem{Biswas:2013xva} 
  S.~Biswas, E.~Gabrielli, F.~Margaroli and B.~Mele,
  JHEP {\bf 07}, 073 (2013)
  [arXiv:1304.1822 [hep-ph]].
\bibitem{Englert:2014pja} 
  C.~Englert and E.~Re, Phys. Rev. D 89, 073020 (2014),
  arXiv:1402.0445 [hep-ph].
\bibitem{Chang:2014rfa} 
  J.~Chang, K.~Cheung, J.~S.~Lee and C.~-T.~Lu, J. High Energy Phys. 05 (2014) 062,
  arXiv:1403.2053 [hep-ph].
\end{thebibliography}
\end{document}